# PRECISION LUMINOSITY MEASUREMENTS AT ILC

I. Bozovic-Jelisavcic[*1], S. Lukic[*], M. Pandurovic[*], I. Smiljanic[*]

*Vinca Institute of Nuclear Sciences, University of Belgrade, Serbia
[1]speaker

[on behalf of the FCAL Collaboration]

**Abstract**

In these proceedings a novel approach to deal with the beam-induced effects in luminosity measurement is presented. Based on the relativistic kinematics of the collision frame of the Bhabha process, the beam-beam related uncertainties can be reduced to the permille level independently of a precision with which the beam parameters are known. Specific event selection combined with the corrective methods we introduce, leads to the systematic uncertainty from the beam-induced effects to be at a few permille level in the peak region above the 80% of the nominal centre-of-mass energies at ILC.

# 1. Introduction

Integrated luminosity measurement at a future linear collider will be performed by counting Bhabha events reconstructed in the luminometer fiducial volume. To match the physics benchmarks (i.e. W-pair production, fermion pair-production, cross-section measurements) that might be of particular interest for the new physics, luminosity should be known at the level of $10^{-3}$ or better [1]. For that purpose, finely granulated calorimeters of high energy and polar angle resolution are foreseen to instrument the very forward region at ILC. Luminometer at ILC will be designed as a compact silicon-tungsten sampling calorimeter with the Moliere radius of approximately 1.5 cm [2].

With the rising energy and the bunch density, one of the main uncertainties in luminosity measurement at a future $e^+e^-$ collider at TeV energies (ILC, CLIC) comes from the effects induced by space charges of the opposite beams. Beamstrahlung (BS) and electromagnetic deflection (EMD) induced by the field of the opposite bunch, together with the initial state radiation (ISR), result in the change of the four-vectors of the initial and final state particles, consequently causing the deviation of the polar angles and counting losses of the signal in the luminometer. Dominating beamstrahlung effects are particularly pronounced at the higher centre-of-mass energies, resulting in 12.8% counting loss at 500 GeV at ILC and 18% counting loss in the 3 TeV CLIC case, in the upper 20% of the luminosity spectrum [3]. At 250 GeV beam-induced effects are smaller, resulting in 8.4% uncertainty in luminosity in the upper 20% of the luminosity spectrum.

# 2. Luminometer at ILD

Two concepts of particle detectors are being developed for ILC, the International Large Detector (ILD) [1] featuring a Time Projection Chamber as the central tracking system, and the Silicon Detector (SiD) [5] with a compact semiconductor central tracker designed to optimize the physics performance as well as the cost. Both detectors have similar layout, combining excellent tracking and finely-grained calorimetry that enables energy reconstruction of individual particles using particle flow algorithm [6]. Design of the luminometer in both detector concepts is the same, the only difference being in the relative longitudinal positioning of the luminometer with respect to the interaction point. The presented study refers to the ILD geometry.

The luminometer itself is foreseen as a sampling silicon-tungsten calorimeter, consisting of 30 absorber planes, each with thickness of one radiation length (3.5 mm), interspersed by segmented silicon sensor planes. To keep the Moliere radius of 1.5 cm, sensor gaps are kept at 1 mm. The accuracy of the electron polar angle reconstruction depends on the sensor segmentation. The optimized layout contains 48 azimuthal and 64 radial divisions, yielding a predicted angular resolution of $\sigma_\theta = (2.20 \pm 0.01) \cdot 10^{-2}$ mrad. This uncertainty contributes to the relative uncertainty of the measured luminosity as $1.6 \cdot 10^{-4}$ [2].

In the ILD version, the luminometer is positioned at 2.5 m from the interaction point, with the geometrical aperture between 31 mrad and 78 mrad and the fiducial volume between 41mrad and 67 mrad. Since the cross-section for Bhabha scattering is falling with the polar angle as $1/\theta^3$, the inner aperture of the luminometer has to be known to 40 μm [9] to keep the counting uncertainty at the same level.

## 3. Beam-induced effects

### 3.1 Beamstrahlung and initial state radiation

As said in Section 1, beamstrahlung emission is the main source of polar angle distortion of the final state electrons and positrons. Bhabha coincidence is lost due to asymmetric boost of one of the final state particles towards the larger polar angles (Figure 1, left).

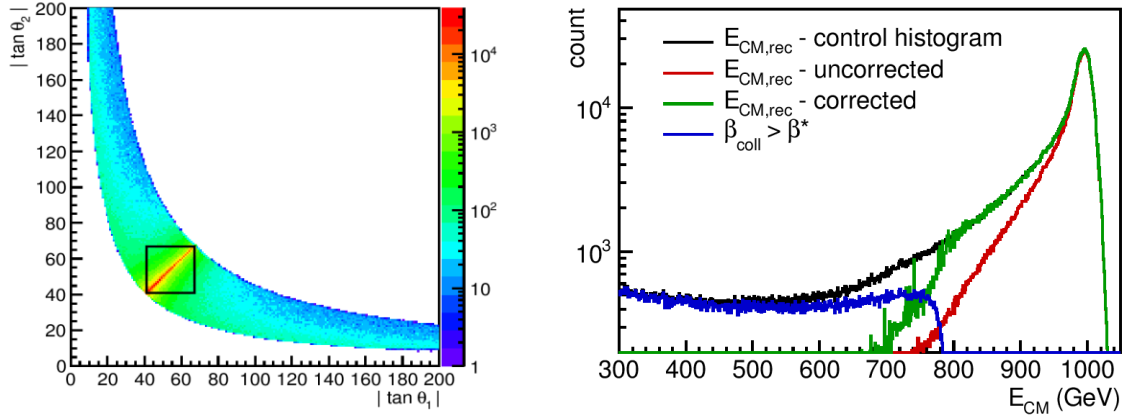

Figure 1: Asymmetric boost of one of the final state electrons (positrons) towards larger polar angles $\theta$ (left); Correction of the signal counting loss due to beamstrahlung and ISR at 1 TeV ILC (right).

The effect can be mathematically described by the fact that the centre-of-mass of the electron-positron system after BS and ISR emission and before emission of the final state radiation (*collision frame*) is moving with a relativistic velocity $\beta_{coll}$ with respect to the laboratory frame. Since the kinematics of the detected showers corresponds to the system after ISR, the velocity of the collision frame $\vec{\beta}_{coll}$ can be calculated from the measured polar angles. As the beamstrahlung and initial state radiation are approximately collinear with the incoming electrons, $\vec{\beta}_{coll}$ can be taken to be approximately collinear with the z-axis. Under this approximation, the modulus of $\vec{\beta}_{coll}$ can be expressed as a function of the measured polar angles of Bhabha partices ($\theta_1^{lab}$ and $\theta_2^{lab}$):

$$\beta_{coll} = \frac{sin(\theta_1^{lab} + \theta_2^{lab})}{sin(\theta_1^{lab}) + sin(\theta_2^{lab})} \quad (3.1)$$

This allows calculating the event-by-event weighting factor $w(\beta_{coll})$ to compensate for the loss of acceptance:

$$w(\beta_{coll}) = \frac{\int_{\theta_{min}}^{\theta_{max}} \frac{d\sigma}{d\theta} d\theta}{\int_{\theta_{min}^{coll}}^{\theta_{max}^{coll}} \frac{d\sigma}{d\theta} d\theta} \quad (3.2)$$

For $\beta_{coll}$ larger than some critical value $\beta^*$ (approximately 0.24 in the ILC case), the effective angular acceptance is zero and such events are inevitably lost (Figure 1, right, blue line). In all other cases, the counting loss can be corrected using Equation 3.2. Correction of the signal counting loss due to the beamstrahlung and ISR is given in Figure 1 (right) for 1 TeV ILC. As can be seen from Figure 1, right (green line), despite severe counting losses due to the beamstrahlung and ISR, the agreement after correction is excellent above the 80% of the nominal centre-of-mass energy. The range below the 80% of the nominal centre-of-mass energy is dominated by the events for which $\beta_{coll}$ is larger than $\beta^*$. The fraction of these events is approximately 14.5%, 15.6% and 17% at 250 GeV, 500 GeV and 1 TeV ILC. However, a small number of events with apparent $\beta_{coll} > \beta^*$ is present at energies above the 80% of the nominal centre-of-mass energy, because the assumption that $\vec{\beta}_{coll}$ is collinear with the beam axis is occasionally broken due to the off-axis initial state radiation. The relative bias due to the off-axis ISR is less than $1.5 \cdot 10^{-3}$ at ILC energies and can be reduced further by appropriate event selection.

The following event selection is applied: Energy of reconstructed Bhabha pair should be larger than 80% of the centre-of-mass energy and the acoplanarity of the reconstructed particles is required to be below 5 deg. Restriction on acoplanarity suppresses events with off-axis radiation to approximately 0.4 permille at all ILC energies. The proposed event selection is not only to minimize the BS- and ISR-related effects but also to suppress background from physics processes (4-fermion production via neutral currents). This is elaborated in detail in [7, 8].

The method presented is data-driven and relies on the experimentally measurable observables - polar angles of the reconstructed Bhabha particles. If the fraction of the lost events in the upper 20% of the luminosity spectrum is taken as a correction, the corresponding uncertainty of the measured luminosity can be reduced further, leaving the remaining average fractional difference due to systematic effects of the method well below a permille [3].

## 3.2 Electromagnetic deflection

Electromagnetic field of the opposite bunch is causing shift of the outgoing particles toward the smaller polar angles. This shift is rather small, but since the Bhabha cross section is monotonously decreasing with the polar angle, the net-effect of the electromagnetic deflection is a decrease of the Bhabha count. This effect is equivalent to the parallel shift in the angular acceptance of the detector fiducial volume by an effective mean deflection angle $\Delta\theta$. Fractional difference in counts ($x_{EMD}$) in the shifted volume with respect to the nominal one can be determined from data (Equation 3.3) and, consequently, the corresponding uncertainty of the measured luminosity $\Delta L_{EMD}$ can be obtained by knowing $\Delta\theta$ (Equation 3.4).

$$x_{EMD} = \frac{1}{N} \frac{dN}{d\theta} \quad (3.3)$$

$$\frac{\Delta L_{EMD}}{L} = x_{EMD}\Delta\theta \tag{3.4}$$

As said before, fractional difference $x_{EMD}$ is a quantity directly accessible in the analysis of either experimental or simulated data, whereas to estimate $\Delta\theta$ beam-beam simulation has to be employed. The following values of $\Delta\theta$ are obtained by simulation: 0.020 mrad at 1 TeV, 0.043 mrad at 500 GeV and 0.067 mrad at 250 GeV. At all ILC energies, $\Delta L_{EMD}/L$ below 5 permille is obtained and can be taken as correction. Systematic uncertainty of this correction comes from the beam parameter variations and it is estimated to be 0.2 permille at 1 TeV, 0.5 permille at 500 GeV and 250 GeV ILC. The method is insensitive to the beam parameter variations up to the 20% variation of bunch charges and sizes [3, 7].

**3.3 Simulation of the signal influenced by the beam-induced effects**

To simulate the influence of the beam-induced effects on signal, Guinea-Pig software 1.4.4 [10] was used. At the point when the initial four-momenta of the colliding electron-positron pairs are generated, the decision is made by Guinea-Pig whether the Bhabha scattering will occur in the collision. The decision is made randomly, based on the cross section for the Bhabha scattering at the centre-of-mass energy of the colliding pair. If Bhabha event is to be realized, the final four-momenta are picked from a file generated with the BHLUMI V4.04 generator [11] at the nominal ILC centre-of-mass energy (500 GeV, 1 TeV). After the event generation, post-generator cuts were applied on the scattering angle $\theta_{coll}$ in the collision frame and only events with 37 mrad $< \theta_{coll} <$ 75 mrad were kept in the event file to be read by Guinea-Pig. The cuts on the scattering angle are somewhat wider than the angular range of the luminometer fiducial volume, to ensure a safety margin for the angular shift due to EMD and the off-axis radiation. The momenta from the generator file are then scaled to the centre-of-mass energy of the colliding pair, rotated to match the collision axis and then boosted back to the laboratory frame. Finally, electromagnetic deflection of the final state is simulated using the Guinea-Pig feature to predict the final deflection angles. The standard beam-parameter set from the ILC Technical Progress Report 2011 [4] is assumed.

**4. Conclusion**

Beam-induced effects at ILC are severe (~10%) if one wants to determine integral luminosity with a permille precision. The dominant counting loss comes from the beamstrahlung and it can be reduced in a simulation-independent manner with a residual uncertainty below a permille. Electromagnetic deflection increases at the lower centre-of-mass energies, up to ~4.3 permille at 250 GeV ILC. From simulation, EMD-induced counting bias can be corrected to better than half a permille at ILC energies.

It is shown that the luminosity can be measured at ILC in the peak region above the 80% of the nominal centre-of-mass energy with the precision of a few permille, even if one corrects for the beam-induced effects without using simulation-dependent information.